\documentclass[12pt,fleqn]{article}
\usepackage{graphics,graphicx,cite}
\usepackage{amssymb,amsfonts}

\def\noi{\noindent}

\def\cm{\hspace*{1cm}}

\def\Jl#1#2{{\it #1\/} {\bf #2},\ }

\def\ApJ#1 {\Jl{Astroph. J.}{#1}}
\def\CQG#1 {\Jl{Class. Quantum Grav.}{#1}}
\def\DAN#1 {\Jl{Dokl. AN SSSR}{#1}}
\def\GC#1 {\Jl{Grav. \& Cosmol.}{#1}}
\def\GRG#1 {\Jl{Gen. Rel. Grav.}{#1}}
\def\JETF#1 {\Jl{Zh. Eksp. Teor. Fiz.}{#1}}
\def\JETP#1 {\Jl{Sov. Phys. JETP}{#1}}
\def\JHEP#1 {\Jl{JHEP}{#1}}
\def\JMP#1 {\Jl{J. Math. Phys.}{#1}}
\def\NPB#1 {\Jl{Nucl. Phys.}{B\ #1}}
\def\NP#1 {\Jl{Nucl. Phys.}{#1}}
\def\PLA#1 {\Jl{Phys. Lett.}{#1A}}
\def\PLB#1 {\Jl{Phys. Lett.}{#1B}}
\def\PRD#1 {\Jl{Phys. Rev.}{D\ #1}}
\def\PRL#1 {\Jl{Phys. Rev. Lett.}{#1}}

\def\eqs{Eqs.\,}
\def\beq{\begin{equation}}
\def\eeq{\end{equation}}
\def\bear{\begin{eqnarray}}

\def\ear{\end{eqnarray}}

\def\nn{\nonumber\\ {}}

\def\eql{&\! = &\!}

\def\tst{\textstyle}

\def\fract#1#2{{\tst\frac{#1}{#2}}}

\def\half{{\fract{1}{2}}}

\def\e{{\,\rm e}}

\def\const{{\rm const}}
\def\eps{\varepsilon}

\def\mn{_{\mu\nu}}

\def\bh{black hole}
\def\bhs{black holes}

\def\ssph{static, spherically symmetric}

\def\ep{\epsilon}

\begin{document}

\title{Cold Black Holes in the Einstein-Scalar Field System}

\author{K.A. Bronnikov\footnote{E-mail: kb20@yandex.ru}\\
Center for Gravitation and Fundamental Metrology, VNIIMS,\\
46 Ozyornaya St., Moscow, Russia;\\
     Institute of Gravitation and Cosmology, PFUR,\\
     6 Miklukho-Maklaya St., Moscow 117198, Russia\\
\\
J.C. Fabris\footnote{E-mail: fabris@cce.ufes.br}\\
Departamento de F\'{\i}sica, Universidade Federal do Esp\'{\i}rito Santo,\\
      Vit\'oria, 29060-900, Esp\'{\i}rito Santo, Brazil\\
\\
N. Pinto-Neto\footnote{E-mail: nelsonpn@cbpf.br}
and M.E. Rodrigues\footnote{E-mail: esialg@cbpf.br}\\
ICRA-CBPF, Rua Xavier Sigaud, 150, Urca, CEP22290-180,\\
Rio de Janeiro, Brazil.}

\maketitle
\begin{abstract}
We study Einstein gravity coupled to a massless scalar field in a
static spherically symmetric space-time in four dimensions. Black
hole solutions exist when the kinetic energy of the scalar field
is negative, that is, for a phantom field. These ``scalar black
holes'' have an infinite horizon area and zero temperature. They
are related through a conformal transformation with similar
objects in the Jordan frames of scalar-tensor theories of gravity.
The thermodynamical properties of these solutions are discussed.
It is proved that any \ssph\ \bhs\ with an infinite horizon area
have zero Hawking temperature.


\end{abstract}

\newpage
%%%%%%%%%%%%%%%%%%%%%%%%%%%%%%%%%%%%%%%%%%%%%%%%%%%%%%%%%%%%%%%%%%%%%%%%%

\section{Introduction}

General relativity predicts the existence of peculiar objects which have
been called black holes \cite{misner}.
Black holes emerge as solutions of Einstein's equations in a static,
spherically symmetric space-time without matter: it is the simplest \bh\
solution found by Schwarzschild in 1916. Its generalization to rotating,
axially symmetric space-times, obtained by Kerr in 1963, also leads to a
black hole structure.
The conventional idea of a black hole implies a singularity in space-time
covered by a horizon. A horizon is a hypersurface which separates an
external region (containing spatial infinity) from an internal region,
which contains a singularity: this internal region is not visible to an
external observer. Generalizations of black hole solutions in different
contexts are known, such as in the presence of an electric field,
non-linear gravity theories, scalar-tensor theories, etc. Their different
properties rise the question of an extension of the \bh\ notion itself.

An example of ``exotic'' black holes are the so-called ``cold black holes''.
They are obtained in scalar-tensor theories in general, and in the
Brans-Dicke theory in particular \cite{lousto,k1,k2}. In general
scalar-tensor theories, the coupling of the scalar field to gravity is
described in terms of a function of the scalar field itself, $\omega(\phi)$,
while in the Brans-Dicke theory this function is simply a constant. From
here on, we will concentrate on the cold black holes that emerge in the
Brans-Dicke theory. It turns out that static, spherically symmetric
solutions of the Brans-Dicke theory reveal a large class of objects that can
be called ``black holes''. Not all of them exhibit a singularity beyond
a horizon: in some cases, the \bh\ interior is regular. However, the horizon
surface has in all such cases an infinite area. Moreover, all such
horizons have zero surface gravity.  This indicates that they have zero
temperature, since the Hawking temperature is directly related to surface
gravity. For this reason, they have been named ``cold black holes''.

The Brans-Dicke theory can be re-defined using a conformal transformation.
By appropriately choosing the conformal factor, the non-minimal coupling
between the scalar field and the scalar curvature, which is an essential
feature of the Brans-Dicke theory, can be broken. In the vacuum case, that
is, with no matter, this results in Einstein gravity minimally coupled to a
massless scalar field.
The energy associated with the scalar field is positive (its kinetic terms
has its usual, ``normal'' sign) if $\omega > - 3/2$; however, the energy is
negative if $\omega < - 3/2$. In this last case, the kinetic term has a
``wrong'' sign, and the theory is called anomalous, or phantom. Such kind of
theories have recently become quite fashionable for both theoretical and
observational reasons. The theoretical reasons are connected with the
ghost condensation and tachyonic fields that result from string theories
\cite{piazza,bagla}.  From the observational viewpoint, recent analysis of
the type Ia supernova data indicates that perhaps the best fit is given by
{\it phantom fields} \cite{caldwell,hannestad,alam,allen}, of which a scalar
field with the ``wrong'' sign of the kinetic term is an example.

The goal of the present work is to show that this Einstein-scalar field
system, with a massless scalar field minimally coupled to gravity, admits
black hole solutions, as opposed to what has been believed
\cite{xanthopoulos}. The black hole solutions can only occur if the energy
associated with the scalar field is negative, that is, for a phantom field.
These ``scalar'' black holes have also an infinite horizon area and zero
temperature. These solutions can be interpreted as Einstein-frame solutions
of the Brans-Dicke theory and transformed to the Jordan frame by performing
the inverse conformal transformation. An interesting point in doing so
is the non-existence of a one-to-one correspondence between black holes
in the Einstein frame and those in the Jordan frame. In this connection, we
are also going to discuss the question of invariance of thermodynamical
quantities, such as the temperature, under these conformal transformations.

In the next section, we reproduce the static, spherically symmetric
solutions for the Einstein-scalar field system. We pay special attention to
a particular class of these solutions. In section 3, we
select which subclass of those solutions corresponds to black holes. In
section 4, we discuss the thermodynamical properties of these objects and
their invariance with respect to conformal transformations. In section 5,
we present our conclusions.

It must be said that Prof. Jos\'e Pl\'{\i}nio Baptista has always had the
Brans-Dicke theory as one of his main interests. Hence, it is a pleasure to
exhibit, in honour of his $70^{\rm th}$ birthday, a new class of black
hole solutions which are somehow related to the Brans-Dicke theory.

\section{Static, spherically symmetric solutions in Einstein-scalar and
    Brans-Dicke theories}

Let us consider gravity coupled minimally to a free massless scalar field.
The action is
\beq
       A = \int d^4x\biggr(R + \epsilon\phi_{;\rho}\phi^{;\rho}\biggl) ,
\eeq
 where $\epsilon = \pm 1$, $+1$ means a normal scalar field with
 positive energy density and $-1$ an anomalous (phantom) scalar field with
 negative energy density.  The field equations are
\bear
       R_{\mu\nu} &=& - \epsilon\phi_{;\mu}\phi_{;\nu} \quad , \\
            \Box\phi &=& 0 \quad .
\ear

The static, spherically symmetric metric may be written as
\beq                                                \label{metric}
        ds^2 = e^{2\gamma}dt^2 - e^{2\alpha}du^2 - e^{2\beta}d\Omega^2 \quad,
\eeq
where $\gamma = \gamma(u)$, $\alpha = \alpha(u)$ and $\beta = \beta(u)$,
$u$ is an arbitrary radial coordinate. Hence, the equations of motion are
\bear                           \label{e1}
    \gamma'' + \gamma'(\gamma' - \alpha' + 2\beta') &=& 0 \quad ,
\\      \label{e2}
     \gamma'' + 2\beta'' + \gamma'^2 + 2\beta'^2 - \alpha'(\gamma' +
    2\beta') &=& - \epsilon\frac{\phi'^2}{2} \quad ,
\\      \label{e3}
    \beta'' + \beta'(\gamma' - \alpha'+ 2\beta') &=& 0 \quad ,
\\          \label{e4}
    \bigr(e^{\gamma - \alpha + 2\beta}\phi'\bigl)' &=& 0 \quad .
\ear
Primes denote derivatives with respect to $u$. The solutions to these
equations have been found in Refs.\,\cite{fisher} for $\ep=1$, \cite{Afish}
for $\ep=-1$ and, in a more general form, \cite{br73}.
They can be obtained by fixing the harmonic radial coordinate, corresponding to
the coordinate condition $\alpha = 2\beta + \gamma$ \cite{br73}. In this
case, the solutions for $\gamma$ and $\phi$ are straightforward,
\beq
    \gamma = - bu, \cm \phi = Cu + \phi_0, \cm b,C,\phi_0 = \const.
\eeq
Using (\ref{e2}) and (\ref{e1}), we obtain the equation
\beq
    f'' + 2bf' +\frac{\epsilon C^2}{2}f = 0, \cm   f = e^{-\beta}
\eeq
with the general solution
\beq
    f = e^{- bu}\{Ae^{ku} + Be^{-ku}\} ,
\eeq
where $k,\ A,\ B$ are integration constants and $2k^2 = 2b^2 -\epsilon C^2$.
The constants $A$ and $B$ obey the relation $4A Bk^2 + 1 = 0$. Requiring that
the solutions should be asymptotically flat at spatial infinity and choosing
$u=0$ for this infinity, we find that the constants must be $A = 1/(2k)$ and
$B = - 1/(2k)$.  Hence the solution takes the form
\beq
    e^\gamma = e^{- bu} , \cm
    e^{\alpha} = \frac{k^2e^{bu}}{\sinh^2 ku}, \cm
    e^\beta = \frac{ke^{bu}}{\sinh ku} .
\eeq

If the parameter $k$ is real and positive, it is helpful to pass over to the
so-called quasiglobal coordinate $\rho$ (such that in (\ref{metric}) $\alpha
+ \gamma =0$) by the transformation
\beq
     \e^{-2ku} = 1 - 2k/\rho \equiv P(\rho).               \label{def_P}
\eeq
The solution takes quite a simple form,
\bear                                                 \label{metricE}
       ds^2_E \eql P^a dt^2 - P^{-a} d\rho^2 - P^{1 - a}\rho^2 d\Omega^2,
\\                          \label{sf}
       \phi \eql -\frac{C}{2k} \ln P(\rho),
\ear
    with the constants related by
\beq                                                        \label{int_0+}
      a = b/k, \cm  a^2 = 1 -\eps C^2/2.
\eeq

In Refs.\,\cite{k1,k2}, the authors were interested in
static, spherically symmetric solutions in the context of scalar-tensor
theories, where the scalar field is non-minimally coupled to gravity,
i.e., in the Jordan frame.
Black hole solutions were also revealed there in the Jordan frame only. Here
we wish to pay attention to the existence of black hole solutions in the
(anti-)Fisher family, or, which is the same, in the Einstein frame of
scalar-tensor gravity, i.e., among the solutions (\ref{metricE}), (\ref{sf}).

The corresponding solutions in the Jordan frame of the Brans-Dicke
theory with the coupling constant $\omega$ may be written in the form
\bear                   \label{metricJ}
    ds^2_J \eql P^\xi\,ds^2_E
\nn
    \eql P^{a - \xi}dt^2 - P^{-a - \xi}d\rho^2 - P^{1 -\xi -a}\rho^2 d\Omega^2,
\\                                                 \label{phiBD}
      \phi_{\rm BD} \eql P^{\xi},
\ear
where $\phi_{\rm BD}$ is the Brans-Dicke scalar related to $\phi$ by
\beq
    \phi_{\rm BD} = \exp \big[\phi/\sqrt{\omega+3/2}\big].
\eeq
The parameters $\xi$ and $a$ are connected by the relation
\beq                                             \label{cond}
    (3 + 2\omega)\xi^2 = 1 - a^2.
\eeq
The subscripts $J$ (Jordan) and $E$ (Einstein) in (\ref{metricE}) and
(\ref{metricJ}) indicate in which frame the solutions are being written.
The two metrics are connected by the conformal transformation
\beq
    g^J_{\mu\nu} = \phi_{\rm BD}^{-1}g^E_{\mu\nu}.
\eeq

\section{Black hole solutions}

Let us try to single out \bh\ solutions from the family (\ref{metricE}),
(\ref{sf}). We restrict ourselves to the the case $k > 0$ since the cases
$k \leq 0$, as may be shown, do not lead to results of equal interest.

First of all, we must fix what we understand by a ``black hole solution''.
For our comparatively simple case of \ssph\ space-times, leaving aside more
general and more rigorous definitions of horizons and \bhs\ (see, e.g.,
\cite{wald}), we can rely on the following working definition. A black hole
is a space-time containing (i) a static region which may be regarded
external (e.g., contains a flat asymptotic), (ii) another region invisible
for an observer at rest residing in the static region, and (iii) a Killing
horizon of nonzero area that separates the two regions and admits an
analytical extension of the metric from one region to another.  This
definition certainly implies that the horizon is regular, since otherwise it
would be a singularity, belonging to the boundary of the space-time
manifold, across which there cannot be a meaningful continuation.

Conditions for having a black hole solution in the Jordan frame, \eqs
(\ref{metricJ}), (\ref{phiBD}), have been discussed in reference \cite{robson}.

For the general metric (\ref{metric}), a \bh\ horizon may be represented by
a sphere $u = u_h$ at which $g_{00} = \e^{2\gamma} =0 $ and at which all
algebraic curvature invariants are finite. To check the latter, it is
sufficient to consider the behaviour of the Kretschmann invariant, given by
\beq
    K = R^{\mu\nu\lambda\gamma}R_{\mu\nu\lambda\gamma}
            = 4K_1^2 + 8K_2^2 + 8K_3^2 + 4K_4^2,
\eeq
where
\bear                                               \label{Kr}
    K_1 &=& {R^{01}}_{01}
        = - e^{-(\alpha + \gamma)}(\gamma'e^{\gamma - \alpha})',
\nn
    K_2 &=& {R^{02}}_{02} = - e^{-2\alpha}\beta'\gamma',
\nn
    K_3 &=& {R^{12}}_{12} = {R^{13}}_{13} =
        - e^{-(\alpha + \beta)}(\beta'e^{\beta-\alpha})',
\nn
    K_4 &=& {R^{23}}_{23} = e^{-2\beta} - e^{-2\alpha}\beta'^2,
\ear
where the primes denote $d/du$.

In the metric (\ref{metricE}), a horizon can occur at $r = 2k$ if $a > 0$.
The radius of the horizon surface is zero (it is thus a center, and a
singularity is suspected) if $1 - a > 0$, infinite if $1 - a < 0$ and finite
if $a = 1$.  The latter case recovers the Schwarzschild solution. Solutions
with $a > 1$ imply an anomalous theory: $\epsilon = - 1$, and
$\omega < - \frac{3}{2}$.

No black holes are thus described by Fisher's solution (\ref{metricE})
in the normal case, $\epsilon = 1$, in agreement with the well-known no-hair
theorems.

Calculating $K_i$ for the metric (\ref{metricE}), we find:
\bear                                                    \label{Kr1}
    K_1 &=& \frac{4ak}{2\rho^3}P^{a - 2}
        \biggr\{1 - (a + 1)\frac{k}{\rho}\biggl\} ,
\nn
    K_2 &=& - \frac{a}{2\rho}P^{a - 2}
        \biggr\{1 - (1 + a)\frac{k}{\rho}\biggl\} ,
\nn
    K_3 &=& - \frac {k}{r\rho^2}P^{a - 2}
        \biggr\{a - (1 + a)\frac{k}{\rho}\biggl\} ,
\nn
    K_4 &=& k\rho P^{a - 2}
        \biggr\{2a - (1 + a)^2\frac{k}{\rho}\biggl\} .
\ear
Therefore, to have regularity at $\rho = 2k$, the condition
\beq
        a - 2 \geq 0                             \label{a>2}
\eeq
must be satisfied. For $a > 2$, the Kretschmann invariant becomes zero.
For $a = 1$ (Schwarzschild's solution) and $a = 2$, it is finite. In other
cases it is infinite, designating a naked singularity. The condition
(\ref{a>2}) is thus necessary for the existence of a horizon and its
regularity. In the Schwarzschild case, the area of the horizon
is finite, given by $S = 16\pi k^2$. In other cases of interest, $a > 2$,
the area of the horizon is infinite. As will be discussed later, this has
important consequences for the thermodynamics of such objects.

However, there is one more condition to be satisfied, namely, that the
geometry should admit an analytical extension beyond the horizon. One may
recall that, for cold black holes in Jordan's frame \cite{k1,k2}, an
analytical extension was found to be only possible for a set of discrete
values of the parameters $\xi$ and $\omega$. Here, the situation is simpler.
A direct inspection of the metric (\ref{metricE}) shows that it is possible
to pass from $\rho > 2k$ to $\rho < 2k$ if the parameter $a$ is an integer, $a
= n$, $n = 2, 3, ...$.  Hence, the objects described by the metric
(\ref{metricE}) are really black holes only if $a = 1, 2, 3,...$ Moreover,
the case $a = 1$ corresponds to the Schwarzschild black hole, for which
$\xi = 0$, the scalar field is constant (actually, absent), and we recover,
as expected, the static, spherically symmetric vacuum solution of Einstein's
equations. A new non-trivial class of black holes is obtained for
\beq   \label{newcond}
        a = 2, 3, ....
\eeq
All these ``anomalous scalar black holes'' (since the scalar field in this
case the scalar field is not constant and contributes to the Einstein
equations with negative energy) have a horizon of infinite area, unlike
the Schwarzschild black hole, in full similarity with the Jordan frame,
but for other values of the solution parameters.

One can reveal one more important difference between the Einstein and Jordan
frames. Namely, as follows from (\ref{Kr1}), all $K_i$ turn to infinity as
$\rho\to 0$ in case (\ref{newcond}). In other words, there is always a
curvature singularity in the internal region of the \bhs\ with a minimally
coupled massless phantom scalar field in general relativity. Meanwhile, many
of Brans-Dicke \bh\ solutions in the Jordan frame are nonsingular, and some
of them have another flat asymptotic beyond the horizon \cite{k1,k2}.

\section{Thermodynamics of the scalar black holes}

Refs.\,\cite{k1,k2} discussed the thermodynamics of scalar-tensor black
holes in the Jordan frame, connected with the presently studied scalar
black holes by conformal transformation. The surface gravity of all these
objects is zero, indicating a zero temperature. The present scalar black
holes share this property.

Indeed, the Hawking temperature is $T_H = (2\pi k_B)^{-1} \kappa$, where
$k_B$ is Boltzmann's constant while the surface gravity $\kappa$ is given
by the expression \cite{wald}
\beq                        \label{kap}
    \kappa = \frac{1}{2}\frac{g'_{00}}
                      {\sqrt{|g_{00}g_{11}|}}\Big|_{u = u_h} ,
\eeq
where $u=u_h$ is the value of the radial coordinate $u$ at the horizon.
After a conformal transformation $g_{\mu\nu} = \Omega^2\tilde g_{\mu\nu}$,
the expression for the surface gravity becomes, in terms of the new metric
$\tilde g_{\mu\nu}$,
\beq                                  \label{conftemp}
    \kappa = \frac{\Omega'}{\Omega}\frac{\tilde g_{00}}
    {\sqrt{\tilde g_{00}\tilde g_{11}}} \Big|_{u = u_h}
        + \frac{1}{2}\frac{\tilde g'_{00}}{\sqrt{\tilde g_{00}
                \tilde g_{11}}}\Big|_{u = u_h}
    = \frac{\Omega'}{\Omega}\frac{\tilde g_{00}}
        {\sqrt{|\tilde g_{00}\tilde g_{11}|}}\Big|_{u = u_h} + \tilde\kappa,
\eeq
where $\tilde\kappa$ is the surface gravity measured in the space-time
with the metric $\tilde g_{\mu\nu}$. The two surface gravity terms ($\kappa$
and $\tilde\kappa$) are equal if the first term in the right-hand side of
(\ref{conftemp}) is zero. We could expect this because the metric component
$g_{00}$ is zero over the horizon. But, we can only assure that $\kappa =
\tilde\kappa$ if the conformal factor $\Omega$ is regular at the horizon,
i.e., if the conformal transformation is well defined on it.

Now, if $g\mn$ is the Jordan-frame metric and ${\tilde g}\mn$ is the
Einstein-frame metric, from  (\ref{metricE}), (\ref{sf}) we
obtain for the temperature in the Einstein frame, with $\Omega =
\phi^{-1/2}$,
\beq
        \tilde\kappa = \frac{a}{2}P^{a - 1}P'\Big|_{\rho=\rho_h} = 0 .
\eeq
Moreover, the surface gravity in Jordan's frame is
\beq
    \kappa = - \frac{\xi}{2}P^{a - 1}P'\Big|_{\rho=\rho_h} + \tilde\kappa = 0,
\eeq
when $a \geq 2$. Hence, the temperature is zero in both frames for the cold
black holes. The conformal factor is regular across the horizon. It may be
tempting to conclude that the \bh\ temperature is conformally invariant
in the case studied here.

The problem of conformal invariance of the Hawking temperature has been
addressed in Ref.\,\cite{jacobson}. In this work, it has been stated
that the Hawking temperature is the same for black holes obtained from
theories which are connected by a conformal transformation under the
conditions of staticity and asymptotic flatness. The situation described
above seems to confirm these general results, but the situation is a little
bit more subtle: it turns out that the very question of conformal
invariance here loses its meaning due to the discrete nature of the \bh\
solutions under consideration.

Indeed, it was established \cite{k1,k2} that black hole solutions, in the
Jordan frame, with $\omega$ constant, that is, in the Brans-Dicke theory,
exist only when the parameters $a$ and $\xi$ obey the relations
\beq
     a = \frac{m + 1}{m - n} , \cm\ \xi = \frac{m - n - 1}{m - n},
\eeq
where $m$ and $n$ are positive integers satisfying the condition
\beq
    m - 2 \geq n \geq 0 .
\eeq
In the Einstein-scalar field system, \bh\ solutions exist under another
condition, (\ref{newcond}). Although the two metrics are connected by the
conformal transformation $g_{\mu\nu} = \phi_{\rm BD}^{- 1}\tilde g_{\mu\nu}$,
the black hole solutions in Jordan's frame do not correspond to black hole
solutions in the Einstein frame, and vice versa. So the question of
invariance loses its meaning here.

There is, however, another, quite general result: {\it any horizon of
infinite area has zero surface gravity (and hence zero Hawking temperature).}
Let us prove it for arbitrary \ssph\ space-times.

For the general form (\ref{metric}) of the metric, the surface gravity
(\ref{kap}) may be written as
\beq
    \kappa = \e^{\gamma-\alpha} |\gamma'| = \half A'(\rho),
\eeq
where $A(\rho) := \e^{2\gamma}$ and $\rho$ is the quasiglobal coordinate
defined by the condition $\alpha + \gamma = 0$. A horizon is a sphere where
$A =0$. So a horizon with finite surface gravity corresponds to a simple
zero of $A$, with $A' \ne 0$, at some finite value of $\rho$. On the other
hand, the regularity conditions require that all $K_i$ (\ref{Kr}) are finite
at the horizon. In particular, in the same coordinates,
$K_2 = -\half A' \beta'$, hence, with $A' \ne 0$, $|K_2| < \infty$ is only
possible in case $|\beta'| < \infty$, which in turn means that $\beta$ is
finite at finite $\rho$. Recalling that $4\pi r^2 = 4\pi\e^{2\beta}$
is the area of the coordinate sphere, we can conclude that a horizon
with finite temperature can only occur at a sphere of finite radius
$r = \e^\beta$. Hence, a horizon with an infinite area can only have zero
temperature.

The above results confirm this general law: all such \bhs\ are perfectly
``cold''.

Finally, the infinite horizon area of the scalar black holes obtained here
may suggest, according to the well-known relations of \bh\ thermodynamics
\cite{wald}, that they have infinite entropy. However, after a more close
investigation, it has been argued that such \bhs\ must in fact have zero
entropy \cite{zaslavskii}, which is more in agreement with a zero
temperature state.  This means that, for such objects, there is a violation
of the law that relates the \bh\ entropy with the horizon area.

\section{Conclusions}

Scalar-tensor theories, which are in general characterized by a non-minimal
coupling between gravity and the scalar field, predict the existence of
exotic black holes, which have an infinite horizon area and zero Hawking
temperature. In the vacuum case, that is, in the absence of any other matter
field, a conformal transformation maps any scalar-tensor theory from a large
class (the Bergmann-Wagoner class) into general relativity minimally
coupled to a massless scalar field.  It had been thought that no black hole
solution exist in this Einstein-scalar field system. We have shown here that
this statement is not true, and we have exhibited a new class of black hole
solutions. However, the price to be paid for their existence is that the
sign of the kinetic term of the scalar field is ``wrong'', that is, the
scalar field has negative energy. As in the scalar-tensor case, the
``scalar'' black holes have infinite horizon areas and zero temperature.
However, the conditions in the parameter space for the existence of such
black holes are different in Jordan's (non-minimal coupling) and
Einstein's (minimal coupling) conformal frames.

One can add that the Einstein frame is common to the whole class of
scalar-tensor theories, whereas Jordan frames change from theory to theory
together with the nonminimal coupling functions. This means that the
discrete ``quantization'' conditions for the solution parameters, providing
the existence of cold \bhs, will be different in similar solutions of
different theories.

Due to the latter circumstance, a discussion of the invariance of the
Hawking temperature with respect to conformal transformations, connecting
different frames, is meaningless for the present solutions:
these transformations do not map a black
hole to a black hole. On the other hand, there is a general law saying that
horizons of infinite area are always ``cold'', i.e., have zero temperature.
So the invariance properties of their temperature become trivial, even if
such \bhs\ are in a conformal correspondence.

The absence of continuations through surfaces of finite (or even zero)
curvature is a peculiar property of many scalar-tensor solutions, indicating
a special type of space-time singularities: violation of analyticity.
Physical properties of such singularities and their possible regularization
by taking into account more general solutions or quantum corrections may be
of considerable interest.

The Hawking temperature discussed here is expressed in terms of the surface
gravity $\kappa$. To be more rigorous, quantum fields around such black
holes must be considered. This is a delicate point, since all black
holes studied in this work have zero temperature, which is, in principle,
a violation of the third law of thermodynamics. For cold black holes in the
Jordan frame, there are anomalies in the definition of quantum fields,
connected with normalization of quantum modes \cite{glauber}. However,
no complete study in this sense has been performed yet, mainly due to
technical difficulties. It would be of interest to consider this problem
in the context of the ``scalar'' black holes presented in this work.

We can add in conclusion that if the phantom scalar field has a nonzero
potential $V(\phi)$, it can form more diverse \ssph\ self-gravitating
configurations including different types of regular black holes
with both zero and non-zero temperature \cite{bf05}.

\medskip\noi
    {\bf Acknowledgments.}

KB thanks the colleagues from DF-UFES for hospitality. The
    work was supported by CNPq (Brazil). J.C.F and N.P-N thank
    also CAPES/COFECUB (Brazil-France scientific cooperation) for partial financial support.

\end{document}